\documentclass[a4paper, 10pt]{elsarticle}
\usepackage{graphicx}
\usepackage{amsmath} 
\usepackage{amssymb}  
\usepackage{amsfonts}%
\usepackage{multirow}
\usepackage{array}
\usepackage{url}
\usepackage[pdftex, plainpages = false, colorlinks=true, linkcolor=black, citecolor = green!50!blue, urlcolor = blue, filecolor=black, pagebackref=false, hypertexnames=false,  pdfpagelabels]{hyperref}

\newtheorem{defn}{Definition}
\newtheorem{alg}{Algorithm}
\newtheorem{assum}{Assumption}
\newtheorem{rem}{Remark}
\newcommand{\trasp}{\ensuremath{^{\intercal}}}
\begin{document}
\begin{frontmatter}
\title{\LARGE \bf Coalitional model predictive control of an irrigation canal\tnoteref{t1}}
\date{October 7, 2023}
\author[esi]{Filiberto Fele\corref{cor1}}
\ead{ffele@us.es}
\author[esi]{Jos\'e M. Maestre}
\author[delft]{Mehdi Hashemy Shahdany}
\author[esi]{\\David Mu\~noz de la Pe\~na}
\author[esi]{Eduardo F. Camacho}%
\address[esi]{Department of Systems Engineering and Automation, University of Seville, 41092 Seville, Spain}
\address[delft]{Department of Water Management, Delft University of Technology, Stevinweg 1, 2628 CN Delft, The Netherlands}
\tnotetext[t1]{\copyright 2014. This manuscript version is made available under the CC-BY-NC-ND 4.0 license \url{https://creativecommons.org/licenses/by-nc-nd/4.0/}.
The final version can be found at \url{https://doi.org/10.1016/j.jprocont.2014.02.005} (please cite as \cite{Fele_2014JPC}).\\
Financial support from the HYCON2 EU-project within the ICT-FP7, and MEC-Spain DPI2008-05818 and FPI grant is gratefully acknowledged.
}
\cortext[cor1]{Corresponding author}

\begin{abstract}
We present a hierarchical control scheme for large-scale systems whose components can exchange information through a data network. The main goal of the supervisory layer is to find the best compromise between control performance and communicational costs by actively modifying the network topology. The actions taken at the supervisory layer alter the control agents' knowledge of the complete system, and the set of agents with which they can communicate. Each group of linked subsystems, or \textit{coalition}, is independently controlled based on a decentralized model predictive control (MPC) scheme, managed at the bottom layer. Hard constraints on the inputs are imposed, while soft constraints on the states are considered to avoid feasibility issues. The performance of the proposed control scheme is validated on a model of the Dez irrigation canal, implemented on the accurate simulator for water systems SOBEK. Finally, the results are compared with those obtained using a centralized MPC controller.
\end{abstract}
\end{frontmatter}
\section{Introduction}
The progress made in data networks technology and the derived decreasing implementation costs --- to which wireless networks have hugely contributed --- have enabled the application of advanced control techniques in systems where the high cost of communication was a substantial obstacle. In particular, large scale systems related with public infrastructures are now fully within the scope of real-time control engineering, fostering the potential of a positive impact on the fundamental services provided countrywide~\cite{NEGENBORN10BOOK}.\par
This paper deals with water management in irrigation canals, a demanding task which entails finding the right trade-off among different sectors in direct competition (agricultural, municipal, and industrial). Since irrigated agriculture constitutes the largest consumer of freshwater resources, the modernization of canal operational management could drastically improve water conservation efficiency and supply flexibility. Moving in this direction, several advanced control strategies have been proposed over the last decades (see, e.g., the survey \cite{MALATERRE98JIDE} and references therein). In~\cite{CLE04JIDE}, an optimal quadratic criteria is used to adjust the parameters of downstream level feedback controllers. Different classes of controllers are considered, ranging from PI controllers at each gate to a centralized controller. The improvement derived from the communication of control actions among neighboring pools is also investigated. In~\cite{PJVO06TH}, the effectiveness of model predictive control in water systems is studied and compared to classical feedback and feedforward strategies.\par
Among several challenging aspects regarding irrigation systems, geographical distance is one of the most interesting. Water networks are generally very disperse, and often different parts of the system are owned by independent entities, expectedly unwilling to coordinate their control actions unless strictly necessary. Moreover, permanent communication between the various parts of the network can be impractical. Consequently, the use of a traditional centralized control approach is hampered, even when the water network is owned and managed by a single entity. Considering all these factors, distributed control schemes can provide solutions able to satisfy the different actors involved. Thus, irrigation canals have become a popular benchmark to assess the performance of hierarchical and distributed control schemes. In~\cite{SCA09JPC}, a thorough classification of these is given. A survey of centralized and distributed MPC schemes for water systems is provided in~\cite{Negenborn:08c}. In this same work, the potential of the application of a distributed MPC scheme --- based on an augmented Lagrangian formulation --- is investigated with a simulation study on an irrigation canal. Also based on an augmented Lagrangian formulation, the work of~\cite{Fawal1998} describes the decomposition of the receding-horizon optimal control problem for heterogeneous irrigation systems, aimed to reduce the computational complexity and to conform to the system topology. The performance objective considered in~\cite{Fawal1998} accounts for the costs of water pumping and water losses, and the profits from power generation. A two-layer control scheme is proposed in~\cite{ZAFRAJMM10JPC}, where the top layer follows a risk management strategy to cope with unexpected changes in the demand, failures or additional maintenance costs, and the bottom layer optimizes the values of water flows by means of a distributed MPC technique. A section of an irrigation canal located in Spain is considered as case study.\par
The idea behind these solutions is to partition the centralized problem among a given number of local controllers or \emph{agents} (see, e.g., \cite{Siljak1991}). Depending on the degree of dynamic interaction between the subsystems, the controllers are categorized in the literature either as distributed or as decentralized. In the first class the agents need to communicate to coordinate their operations~\cite{NEG1,VEN06TH,JMM10OCAM,Farina:2012:DPC}. By contrast, in the second class the limited degree of interaction allows the agents to tackle their control tasks with no need of communication~\cite{ALE07ECC,MS06Automatica}. Between these two classes lie the \emph{coalitional} control schemes. In these, the control strategy is adapted to the varying coupling conditions between different parts of the system, promoting cooperation among the control agents most concerned at any given time. The formation of groups of cooperative agents based on the active coupling constraints is considered in~\cite{Trodden09}. The work of~\cite{JilgStur2013_IFAC} describes a hierarchical framework where information among the agents is exchanged at each time step within clusters of strongly dynamically coupled subsystems, while a slower communication rate is required between different clusters. In~\cite{Ocampo2012HierarcMPC}, the complexity of the model predictive control problem of the Barcelona drinking water network is reduced by means of a partitioning algorithm, in order to control in a hierarchical-distributed manner the resulting subnetworks. In~\cite{NunezEtAl2013} a flexible hierarchical MPC scheme is proposed for a hydro-power valley, where the priority of the agents in optimizing their control actions can be rearranged according to the different operational conditions.\par
The present work focuses on how the interaction between the subsystems varies with time. Consider a large-scale system with an associated network, in which a number of control agents communicate in order to derive their knowledge of the overall system. Here, a two-layer hierarchical control strategy that manipulates the network topology with regard to both the current state of the system and the communicational cost is proposed. Those data links that do not yield a significant improvement of the control performance, compared with their relative cost of use, are disconnected. This feature is interesting, e.g., for communication infrastructures based on battery-powered wireless communication devices. Thus, any agent will be able to communicate only with those agents whose cooperation is most relevant, and the overall system will be partitioned into \textit{coalitions} working in a decentralized fashion.\par
The properties of a multi-agent control scheme based on this idea are discussed in~\cite{Maestre2013OCAM}, where the time-variant relevance of the communication within a set of dynamically coupled, unconstrained linear systems is analyzed using tools from cooperative game theory~\cite{SAAD09IEEESPM}. The research was extended towards input constrained systems in a preliminary version of this work \cite{FELE13ICSNC}, employing a model predictive control strategy~\cite{EFC1,RawlingsLIB09} at the bottom layer. A local Luenberger observer was used within each coalition in order to estimate the dynamic influence caused by external subsystems.\par
In this article, constraints on both states and inputs are considered, and local Kalman filters are used to estimate the dynamic couplings between different coalitions, viewed as perturbations. Information about any measurable disturbance can be now employed at the bottom layer. The proposed control scheme is validated on a detailed model of a 45 km section of the Dez irrigation canal, implemented on the SOBEK hydrodynamic simulator. For comparison, the results are shown along with those obtained using a centralized MPC controller.\par
The paper is organized as follows. In \S \ref{sec_prob_form}, a formulation of the control problem is presented. The new distributed control algorithm is introduced in \S \ref{sec_dist_contr}. The performance of the controller is finally validated in \S \ref{sec_canalDez}, employing a model of a section of the Dez irrigation canal as case study.
\section{Problem formulation}
\label{sec_prob_form}
\subsection{System model}
On the basis of the work of \cite{PJVO06TH}, where the implementation of model-based control techniques on water systems has been examined, a discrete-time linear approximation of the dynamics of the irrigation canal, namely the ID model~\cite{Schuurmans1997}, is adopted here. According to this model, each reach is characterized as a transport delay in series with an integrator.\footnote{A more detailed description of the model and the parameters of the canal is given in \S \ref{sec_canalDez}.} A minimal order representation of the dynamics of the canal follows, which is suitable for the application to the control scheme presented in this article, where the computational burden is a major issue.\par
The canal is partitioned into its essential components, i.e., pairs composed by a gate (the actuator) and its downstream reach, forming a set $\mathcal{V} = \{1,\ldots,N\}$ of subsystems. The dynamics of any subsystem $i\in\mathcal{V}$ are described by the linear model:
\begin{subequations}\label{eq:system}
\begin{align}
x_i(k+1) &= A_{ii}x_i(k) + B_{ii}u_i(k) + E_i p_i(k) + G_i w_i(k)\label{eq:system_a}\\
w_i(k) &= \sum_{j\in \mathcal{N}_i} A_{ij}x_j(k)+ B_{ij}u_j(k)\label{eq:system_b}%
\end{align}
\end{subequations}
where $x_i \in \mathbb{R}^{n_i}$ and $u_i \in \mathbb{R}^{m_i}$ are the state and input vectors respectively, $p_i \in \mathbb{R}^{l_i}$ is a measurable perturbation due to the offtake flow, and $w_i \in \mathbb R^{r_i}$ describes the influence on $x_i$ of the neighbors' states and inputs. In \eqref{eq:system_b}, $x_j \in \mathbb{R}^{n_j}$ and $u_j \in \mathbb{R}^{m_j}$ are the state and input vectors of each neighbor $j \in \mathcal{N}_i$ of subsystem $i$. The neighborhood set $\mathcal{N}_i$ is defined as:
\begin{equation}\label{eq_neigh_set}
\mathcal{N}_i = \left\{j\in \mathcal{V}| A_{ij} \neq \mathbf{0} \vee B_{ij} \neq \mathbf{0}, j\neq i  \right\}
\end{equation}
i.e., it is the set composed by any subsystem $j\neq i$ whose state and/or input produce some effect on the dynamics of subsystem $i$.
\begin{rem}
Even if model \eqref{eq:system} is tailored to the case study, it is indeed a general formulation fitting a wide variety of large-scale systems.
\end{rem}
The state vector
\[
x_i(k) \equiv [q_i(k-1), \ldots, q_i(k-d_i), e_i(k)]\trasp
\]
gathers information about the flow $q_i$ along the reach and the water level error $e_i$ with respect to a desired value. Notice that an augmented representation is used in order to take into account the flow transport delay $d_i$. The input $u_i(k) \equiv \Delta q_i(k)$ is the variation of the flow entering the reach $i$, controlled at its upstream gate.\par
For each subsystem, the measure of the water level error $e_i(k)$ in the backwater section of the reach is available to its control agent; the rest of the state variables (water flow in different sections of the reach) are observable.
\subsection{Exchange of information}
All the control agents can communicate through a data network whose topology is described by means of the undirected graph $\mathcal{G}=(\mathcal{V},\Lambda)$, where to each subsystem in $\mathcal{V}$ is assigned a node. Let $\mathcal{L} \subseteq \mathcal{V} \times \mathcal{V}$ be the set of edges corresponding to the existing communication links between the agents. Each link $\ell_{ij} = \left\{i,j\right\} = \left\{j,i\right\} = \ell_{ji} \in \mathcal{L}$ can be either enabled or disabled. Then the \emph{network topology} $\Lambda(k) \subseteq \mathcal{L}$ is defined as the set of links enabled at a given time, i.e., $\ell_{ij}\in \Lambda(k)$ if and only if it is enabled at time $k$. Each active link has a cost $c_{\ell}>0$ per time of use. This cost can vary, e.g., as a function of the available bandwidth. For simplicity, a constant and unique value of $c_{\ell}$ is considered in the remainder for all the links.\footnote{The value of $c_{\ell}$ is a given parameter in the described framework; also, it is 
assumed as constant during the prediction horizon. With these premises, the complexity of the topology selection problem is not related to differences in the costs of use of the links. On the other hand, the solutions of the problem will depend on the costs.}
\begin{defn}\label{def2}
Any two agents are said to be \emph{connected} if and only if there exists a path between them in $\mathcal{G}=(\mathcal{V},\Lambda)$.
\end{defn}
\begin{assum}\label{assum1}
Any two agents can communicate if and only if they are connected.
\end{assum}
From Definition \ref{def2} and Assumption \ref{assum1} it follows that any given network topology induces a partition of the whole agent set $\mathcal{V}$ into disjoint communication components~\cite{MSAVDN01}. As agents within the same communication component will benefit from cooperation --- i.e., sharing information in order to aggregate their control tasks --- we will refer to such components as \textit{coalitions}, and the partition resulting by a given network topology $\Lambda(k)$ will be denoted as $\mathcal{V}/\Lambda = \{\mathcal{C}_1,\mathcal{C}_2,\ldots,\mathcal{C}_{|\mathcal{V}/\Lambda|}\}$, where $|\cdot|$ represents the cardinality of the set. To ease the notation, let us define the set of indices $\mathcal{V}_{\Lambda} = \{1,\ldots,|\mathcal{V}/\Lambda|\}$. Any partition $\mathcal{V}/\Lambda$ originates a set of coalitions satisfying the following conditions~\cite{Rahwan2012}:
\begin{enumerate}[(i)]
	\item $\mathcal{C}_i\neq \varnothing$, $\forall i \in \mathcal{V}_{\Lambda}$;
	\item $\bigcup\limits_{i=1}^{|\mathcal{V}/\Lambda|}\mathcal{C}_i = \mathcal{V}$;
	\item $\mathcal{C}_i \cap \mathcal{C}_j = \varnothing$, $\forall i,j\in \mathcal{V}_{\Lambda}$, $i\neq j$.
\end{enumerate}
The number of coalitions $|\mathcal{V}/\Lambda|$ produced by any topology $\Lambda$ pertains to the interval $\left[1,N\right]$, whose extremes represent the centralized control case (all the $N$ subsystems connected\footnote{Notice that, according to Assumption \ref{assum1}, this condition does not necessarily require all the links to be active.}) and the case where each subsystem forms a coalition on its own (all links disabled).
\subsection{Coalition dynamics}
Thus, to describe the dynamics of each coalition $\mathcal{C}_i \in \mathcal{V}/\Lambda$, the following extension of \eqref{eq:system_a} holds:
\begin{equation}\label{eq_coal}
\xi_i(k+1) = \Xi_{ii} \xi_i(k) + \Upsilon_{ii}\upsilon_i(k) + \Phi_i \rho_i(k) + \Psi_i\omega_i(k)
\end{equation}
where $\xi_i$ and $\upsilon_i$ are respectively the state and input vectors of coalition $\mathcal{C}_i$, composed by stacking the vectors of all the subsystems in the coalition:
\begin{equation*}
\xi_i=\{x_s\}_{s\in\mathcal{C}_i}, \quad \upsilon_i=\{u_s\}_{s\in\mathcal{C}_i}, \quad i \in \mathcal{V}_{\Lambda}
\end{equation*}
Similarly, $\rho_i = \{p_s\}_{s\in\mathcal{C}_i}$ is the vector grouping the offtake flows. As an extension of \eqref{eq:system_b}, $\omega_i$ expresses the influence of the neighbor coalitions' states and inputs on $\xi_i$:
\begin{equation}
\omega_i(k)=\sum_{j\in \mathcal{N}_{\mathcal{C}_i}} \Xi_{ij} \xi_j(k)+ \Upsilon_{ij}\upsilon_j(k)
\label{eq_ext_influence}
\end{equation}
where the set of neighbors $\mathcal{N}_{\mathcal{C}_i}$ is an extension of \eqref{eq_neigh_set}, indexing any coalition $\mathcal{C}_j$, $j\neq i$ whose state and/or inputs produce some effect on the dynamics of (any subsystem inside) coalition $\mathcal{C}_i$. Matrices $\Xi_{ii}$, $\Xi_{ij}$, $\Upsilon_{ii}$, $\Upsilon_{ij}$, $\Phi_i$ and $\Psi_i$ are composed accordingly.
\subsection{Control objective}
In the application considered here, the control objective is to regulate the water level error of all the reaches to zero while minimizing a cost that depends on the state and input trajectories. An additional term in the cost function will take into account the use of network resources.\par
In order to meet the objective, the offset caused by the offtake flows is canceled by steering each coalition's state to a suitable setpoint $\left\{\bar{\xi}_i, \bar{\upsilon}_i \right\}$. By imposing the steady state condition for all the coalitions, and setting the water level errors to zero, the following system of equations is obtained:
\begin{equation}
\left[
\begin{array}{cc}
	I-\boldsymbol{\Xi} & -\boldsymbol{\Upsilon}\\
	\boldsymbol{\Gamma} & 0
\end{array}
\right]
\left[
\begin{array}{c}
	\bar{\boldsymbol{\xi}}\\
	\bar{\boldsymbol{\upsilon}}
\end{array}
\right] =
\left[
\begin{array}{c}
	\boldsymbol{\Phi} \boldsymbol{\rho}(k)\\
	0
\end{array}
\right]
\label{eq_setpoint}
\end{equation}
where $\boldsymbol{\xi} \equiv \{\xi_i\}_{i\in\mathcal{V}_{\Lambda}}$ and $\boldsymbol{\upsilon}\equiv \{\upsilon_i\}_{i\in\mathcal{V}_{\Lambda}}$, i.e., the global state and input vectors permuted according to the partition $\mathcal{V}/\Lambda$. Similarly, $\boldsymbol{\Xi}\in \mathbb{R}^{n\times n}$ and $\boldsymbol{\Upsilon}\in \mathbb{R}^{n\times m}$ are the permuted global state and input matrices. The offtake vector $\boldsymbol{\rho} \equiv \{\rho_i\}_{i\in\mathcal{V}_{\Lambda}}$ and $\boldsymbol{\Phi}$ are composed likewise. The matrix $\boldsymbol{\Gamma} \in \mathbb{R}^{N\times n}$ is determined such that $\boldsymbol{\Gamma}\boldsymbol{\xi}$ is the stacked vector of the water level errors of all the coalitions, i.e.:
\begin{equation*}
\boldsymbol{\Gamma}\boldsymbol{\xi} = \{e_s\}_{s\in\mathcal{C}_i},\quad \forall i \in\mathcal{V}_{\Lambda}
\end{equation*}
Denoting the shifted state and input of coalition $\mathcal{C}_i$ as $\zeta_i = \xi_i - \bar{\xi}_i$ and $\nu_i = \upsilon_i - \bar{\upsilon}_i$, respectively, the cost function can be divided into a term $J_s$ representing the optimal performance objective, and a term $J_n$ expressing the network-related cost:
\begin{subequations}\label{eq_func_coste}
\begin{align}
J_{s,i} & = \sum\limits_{t=0}^{N_p-1} \left( \zeta\trasp_i(t|k)Q_i \zeta_i(t|k) + \nu_i\trasp(t|k)R_i \nu_i(t|k)\right) + \zeta\trasp_i(N_p|k)P_i \zeta_i(N_p|k)\label{eq_coste_opt}\\
J_{n,j} & = N_p \frac{c_{\ell}}{2}n_{\ell,j}\left(\Lambda\right) \label{eq_coste_red}
\end{align}
\end{subequations}
where the notation $\alpha(t|k)$ corresponds to the value of $\alpha$ predicted at time $k+t$, based on the knowledge at time $k$, and $Q_i \geq 0$, $R_i > 0$ and $P_i = P_i\trasp > 0$ are constant weighting matrices.\footnote{Matrix $Q_i$ only weights deviations of the water level.} In \eqref{eq_coste_red}, $n_{\ell,j}(\Lambda)$ is the number of active links \emph{directly} connecting agent $j$ to other agents according to the network topology $\Lambda$. Note that each agent shares the cost of a link with the agent located at the other side of that link.
The overall control problem can be posed as the following receding-horizon optimization:
\begin{equation}\label{eq_centr_prob}
\min\limits_{\boldsymbol{\upsilon},\Lambda} \sum\limits_{i\in\mathcal{V}_{\Lambda}} J_{s,i}(\xi_i(k),\upsilon_i,\Lambda) + \sum\limits_{j\in\mathcal{V}} J_{n,j}(\Lambda)
\end{equation}
s.t.
\begin{subequations}\label{eq_centr_prob_constr}
\begin{equation*}
\begin{split}
\xi_i(t+1|k) & = \Xi_{ii} \xi_i(t|k) + \Upsilon_{ii}\upsilon_i(t|k) + \Phi_i \rho_i(k) + \Psi_i \hat{\omega}_i(k)\\
\xi_i(t|k) & \in \mathcal{X}_i,\; \forall t\in \left[0, N_p\right]\\
\upsilon_i(t|k) & \in \mathcal{U}_i,\; \forall t\in \left[0, N_p - 1\right]\\
\xi_i(0|k) & = \xi_i(k)\\
\Lambda & \subseteq \mathcal{L}
\end{split}
\end{equation*}
\end{subequations}
\begin{rem}\label{rem1}
In general, as the planning of the offtakes is made in advance, the perturbation $\rho_i(k + t)$ can be known for the entire prediction horizon. In this work, however, forecasts are not considered, and a constant value of the offtake flow, corresponding to the current measure, is maintained along the horizon.
\end{rem}
\begin{rem}
Notice that \eqref{eq_ext_influence} cannot be used directly, as coalition $\mathcal{C}_i$ has no knowledge of the states and inputs of external subsystems. An estimate $\hat{\omega}_i$ of the perturbation they cause on $\xi_i$ is thus performed at each time step $k$, and its value is assumed constant along the prediction horizon. Details are given in \S\ref{sec_bottom_layer}.
\end{rem}
Problem \eqref{eq_centr_prob} constitutes a dynamic programming optimization with mixed integer variables, which is generally not practical to solve. Notice that the integer variables do not explicitly appear in \eqref{eq_centr_prob}. Since any topology corresponds to a partition of the global system, the composition of the resulting coalitions' state and input vectors and matrices will implicitly depend on $\Lambda$. The choice of the network topology is made within a discrete set whose size, in the general case, grows exponentially with the number of links. In the remainder, we formulate a hierarchical multi-agent control algorithm which provides a suboptimal, yet less computationally expensive solution.
\section{The control algorithm}
\label{sec_dist_contr}
\begin{figure}[tbh]
  \centering
    \includegraphics[width=0.8\columnwidth]{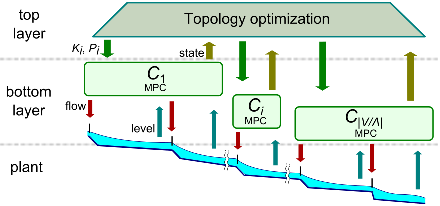}
    \caption{Functional diagram of the coalitional MPC.}
    \label{fig_esquema_func}
\end{figure}

The hierarchical multi-agent strategy proposed in this article is based upon an approximation of problem \eqref{eq_centr_prob}, with the intent of reducing the computational requirements when dealing with large-scale systems. The resulting control problem will provide a suboptimal solution to \eqref{eq_centr_prob}, and it will be split in two layers: a top layer that will take charge of the choice of the network mode, and a bottom layer which will handle the estimation and the real-time control tasks. A conceptual diagram of the proposed coalitional MPC strategy is shown in Figure~\ref{fig_esquema_func}.\par
\subsection{Top layer}
\label{sec_toplayer}
The discrete part of \eqref{eq_centr_prob}, related to the choice of the optimal network topology, is separated and assigned to the top layer. As this constitutes the most computationally demanding part of the problem, its solution is computed on a coarser time scale (with respect to the sample time required for the control of the system). Several network topologies are compared at the top layer, and the most appropriate is selected for the following interval $T_{\Lambda}$. Let $\mathfrak{L}^{+} = \{\Lambda_1,\Lambda_2,\ldots,\Lambda_{|\mathfrak{L}^{+}|}\}$ be the set of possible network topologies to be evaluated. Then, let us define the function $J: \mathbb{R}^{n}\times \mathcal{L} \mapsto \mathbb{R}$ as follows~\cite{MAESTRE11IFAC}:
\begin{equation}\label{eq:v_LG}
J(\boldsymbol{\xi},\Lambda)= \sum_{i\in\mathcal{V}_{\Lambda}}{\zeta_i\trasp P_i \zeta_i} + c_{\ell}|\Lambda|T_{\Lambda},\; \Lambda\in \mathfrak{L}^{+}
\end{equation}
where $P_i = P_i\trasp > 0$, $|\Lambda|$ is the number of enabled links and $c_{\ell}$ is the cost of use of one link, considered over the interval $T_{\Lambda}$.
It is not pragmatic to see $\mathfrak{L}^{+}$ as the set containing every possible configuration of links. Because the number of all possible topologies grows exponentially with the number of links, the set $\mathfrak{L}^{+}$ should be defined as a reasonably sized set of \emph{relevant} topologies for the system to be controlled. The composition of $\mathfrak{L}^{+}$ could either be static or evolving depending, for example, on the current state of the system, on the network constraints, or on the willingness to cooperate among the agents. Of all the configurations considered at a given moment, the one giving the minimum value of \eqref{eq:v_LG}, denoted as $\Lambda^{\ast}\in\mathfrak{L}^{+}$, will be applied during the next interval $T_{\Lambda}$.\par
As a consequence to the choice of any given topology $\Lambda$, the set of agents is partitioned into a specific set of coalitions $\{\mathcal{C}_1,\mathcal{C}_2,\ldots,\mathcal{C}_{|\mathcal{V}/\Lambda|}\}$. To attain the optimal performance objective \eqref{eq_coste_opt}, a feedback gain $\mathbb{K}$ for the whole system is computed at the top layer.\footnote{In the global coalition case, the feedback law $\mathbb{K}$ will coincide with the LQR gain.} In conformity with the system partition $\mathcal{V}/\Lambda$, $\mathbb{K}$ will be composed of a set of decentralized feedback gains, each one associated to a coalition, i.e., $\mathbb{K}=diag\{K_1,\ldots,K_{|\mathcal{V}/\Lambda|}\}$.
Let $\mathbb{P} > 0$ be the block matrix having $\{P_1,\ldots,P_{|\mathcal{V}/\Lambda|}\}$ on its diagonal, and consider the Lyapunov function $V(\boldsymbol{\xi}) = \boldsymbol{\xi}\trasp \mathbb{P} \boldsymbol{\xi}$, where $\boldsymbol{\xi} \equiv \{\xi_i\}_{i\in\mathcal{V}_{\Lambda}}$ is the global state vector, permuted according to the partition $\mathcal{V}/\Lambda$. For $V(\boldsymbol{\xi})$ to constitute an upper bound on the infinite-horizon performance objective, the constraints of the following LMI problem have to be satisfied (see for example \cite{KBM96}):
\begin{equation}\label{eq_sel_mode}
\max_{\mathbb{K},\mathbb{P}}\mathbf{Tr}\mathbb{P}^{-1}
\end{equation}
s.t.
\begin{equation*}
\begin{split}
& \mathbb{P} = \mathbb{P}\trasp > 0\\
& (\mathbf{\Xi} + \mathbf{\Upsilon} \mathbb{K})\trasp \mathbb{P}(\mathbf{\Xi} + \mathbf{\Upsilon} \mathbb{K}) - \mathbb{P} \leq - \mathbf{Q} - \mathbb{K}\trasp \mathbf{R} \mathbb{K}
\end{split}
\end{equation*}
where $\boldsymbol{\Xi}$ and $\boldsymbol{\Upsilon}$ are respectively the global state and input matrices, composed to match $\boldsymbol{\xi}$ and $\boldsymbol{\upsilon}\equiv \{\upsilon_i\}_{i\in\mathcal{V}_{\Lambda}}$. Similarly, $\mathbf{Q} = diag\{Q_1,\ldots,Q_{|\mathcal{V}/\Lambda^{\ast}|}\} \geq 0$ and $\mathbf{R} = diag\{R_1,\ldots,R_{|\mathcal{V}/\Lambda^{\ast}|}\} > 0$ are the global weighting matrices.\par
By the solution of \eqref{eq_sel_mode}, a set of feedback control laws $\upsilon_i = K_i \xi_i$ which minimize $V(\boldsymbol{\xi})$ and a set of matrices $P_i$, $i\in \mathcal{V}_{\Lambda}$ is obtained. These matrices are then used to compute the value of~\eqref{eq:v_LG} and find its minimizer $\Lambda^{\ast}$.
\begin{rem}
Notice that the evaluation of different network topologies is independent and can be executed in parallel on a multi-processor platform. Also, the set of control laws associated with any network topology could be stored and reused whenever the same topology is considered again, without the need of solving more than once the relative LMI problem.
\end{rem}
\subsection{Bottom layer}
\label{sec_bottom_layer}
At the bottom layer, the control is decentralized into the coalitions arising from the partition $\mathcal{V}/\Lambda^{\ast}$.
With the term \emph{decentralized} we designate the complete absence of communication among different coalitions; agents within a coalition share their information at each sample time $k$. As a consequence, the term $\omega_i$ related with the water demand in neighboring coalitions cannot be computed through \eqref{eq_ext_influence}. Every coalition gets an estimate $\hat{\omega}_i$ by means of a local Kalman filter, based on the available measures of water level errors and current offtake flows. Given the slow dynamics of the system and the steady nature of the offtake flows, transient dynamics are neglected. Specifically, $\hat{\omega}_i$ is viewed as a constant integrating disturbance, included in the model defining a suitable augmented state vector.
\begin{rem}
In general, inter-pool transient dynamics are not negligible. However, \emph{(i)} the system examined in the case study is inherently stable, and \emph{(ii)} the presence of local flow controllers at each gate is considered. If the response of these local control loops is sufficiently fast, the interaction between two adjacent pools reduces to a one-way perturbation~\cite{Schuurmans1997}.
\end{rem}
Then, implementing a standard offset-free scheme for regulation~\cite{2003PannocchiaRawlings}, the agents steer their subsystems to an appropriate setpoint in order to compensate for both the estimated disturbance and the offset caused by the measurable offtake flows. Notice that any offset due to mismatches between the linear model and the actual system will be also included in $\hat{\omega}_i$. The Kalman filter also serves as an observer for the values of water flows.\par
For any coalition $\mathcal{C}_i$, the setpoint $\{\bar{\xi}_i,\bar{\upsilon}_i\}$ is obtained by the solution of the linear system:
\begin{equation}\label{eq_setpoint_botlay}
\left[
\begin{array}{cc}
	I-\Xi_{ii} & -\Upsilon_{ii}\\
	\Gamma_i & 0
\end{array}
\right]
\left[
\begin{array}{c}
	\bar{\xi}_i\\
	\bar{\upsilon}_i
\end{array}
\right] =
\left[
\begin{array}{c}
	\Phi_i \rho_i(k) + \Psi_i\hat{\omega}_i(k)\\
	0
\end{array}
\right]
\end{equation}
where $\Gamma_i$ is an output matrix defined such that $\Gamma_i \xi_i = \{e_s\}_{s\in\mathcal{C}_i}$, i.e., the vector composed by the water level deviations of all the subsystems in coalition $\mathcal{C}_i$. Note that, because of the estimation of $\hat{\omega}_i$, the setpoint computed through \eqref{eq_setpoint_botlay} is expected to change at each time step. Moreover, when the topology is changed, the Kalman filters structure changes as well, according to the composition of the new set of coalitions. To avoid undesired drifts on the computed value of the setpoint, a good initial guess of the state and the covariance matrix is needed for each new coalition's local filter. Assuming that each agent is able to communicate past data to any other member of the coalition, the initial guess is obtained by performing a few iterations of the Kalman filter on these past data.\par
The setpoint obtained by \eqref{eq_setpoint_botlay} may not satisfy the constraints. Therefore, the problem~\eqref{eq_setpoint_opt} is minimized to obtain the nearest feasible setpoint $\{\xi_i^s,\upsilon_i^s\}$ to that resulting from \eqref{eq_setpoint_botlay}. This requires the introduction of the slack variable $\sigma_i$ in the equality constraint.
\begin{equation}\label{eq_setpoint_opt}
\min\limits_{\xi^s_i,\upsilon^s_i,\sigma_i} (\upsilon_i^s-\bar{\upsilon}_i)\trasp R_i (\upsilon^s_i-\bar{\upsilon}_i) + {\xi_i^s}\trasp Q_i \xi_i^s + \sigma_i\trasp G_i \sigma_i
\end{equation}
s.t.
\begin{equation*}
\begin{split}
& (I-\Xi_{ii})\xi^s_i -\Upsilon_{ii}\upsilon^s_i -\sigma_i = \Phi_i \rho_i(k) + \Psi_i\hat{\omega}_i(k)\\
& \Phi_i \xi_i^s > 0\\
& K_i(\xi_i(k) - \xi^s_i) + \upsilon^s_i \in \mathcal{U}_i
\end{split}
\end{equation*}
In~\eqref{eq_setpoint_opt}, the input reference $\bar{\upsilon}_i$ is the one obtained from \eqref{eq_setpoint_botlay}, $Q_i$ and $R_i$ are the same used for the optimal performance specification in the top layer, and $G_i > 0$ is a constant weighting matrix, whose value has been chosen such that the magnitude of the term $\sigma_i\trasp G_i\sigma_i$ in~\eqref{eq_setpoint_opt} is comparable with the rest of the cost function. Notice that hard constraints are imposed here on the water flows: the product $\Phi_i \xi_i^s$ is the vector composed of the flow values at each section of any reach controlled by coalition $\mathcal{C}_i$. Finally, the shifted state is redefined as $\zeta_i = \xi_i - \xi_i^s$.\par
Based on this shifted state, the following controller
\begin{equation}
\upsilon_i(k) = K_i\zeta_i(k) + \upsilon_i^s(k)
\label{eq_input_noMPC}
\end{equation}
is available for each coalition to regulate the subsystems to the desired setpoint. However, feasibility cannot be guaranteed with the control law \eqref{eq_input_noMPC}. Thus, the value given by \eqref{eq_input_noMPC} is ``rectified'' through the solution of an MPC problem, obtaining an additional input term:
\begin{equation}
\upsilon_i(k) = K_i\zeta_i(k) + \upsilon_i^s(k) + \upsilon'_i(k)
\label{eq_input_MPC}
\end{equation}
Possible issues related with the loss of feasibility are dealt with by considering physical limits on the water flows as soft constraints. Restrictions on the input change rate are formulated as hard constraints. The shifted input can be now redefined as $\nu_i = \upsilon_i - \upsilon_i^s$. The cost function is derived from \eqref{eq_coste_opt} as follows:
\begin{multline}\label{eq_coste_MPC}
J'_{i} = \sum\limits_{t=0}^{N_p-1} \left( \zeta\trasp_i(t|k)Q_i \zeta_i(t|k) + \nu_i\trasp(t|k)R_i \nu_i(t|k)\right) + \\
 + \zeta\trasp_i(N_p|k)P_i \zeta_i(N_p|k) + \sum\limits_{t=1}^{N_p} \epsilon_i\trasp (t)S_i\epsilon_i(t)
\end{multline}
where $S_i > 0$ is a constant weighting matrix for the slack variable $\epsilon_i$, used to relax the constraints on the flows. The optimization problem to be solved by coalition $\mathcal{C}_i$ at each time step $k$ is:
\begin{equation}\label{eq_decentr_MPC}
\min\limits_{\upsilon'_i,\epsilon_i} J'_{i}(\zeta_i(k),\nu_i)
\end{equation}
s.t.
\begin{subequations}\label{eq_decentr_prob_constr}
\begin{equation*}
\begin{split}
& \zeta_i(t+1|k) = (\Xi_{ii}+ \Upsilon_{ii}K_i)\zeta_i(t|k) + \Upsilon_{ii}\upsilon'_i(t|k)\\[5pt]
& \Phi_i\big(\zeta_i(t|k) + \xi_i^s(k)\big) + \epsilon_i(t) > 0,\; \forall t\in \left[0, N_p\right]\\
& K_i\zeta_i(t|k) + \upsilon'_i(t|k) + \upsilon_i^s(k) \in \mathcal{U}_i,\; \forall t\in \left[0, N_p\right]\\
& \zeta_i(0|k) = \xi_i(k) - \xi_i^s(k)
\end{split}
\end{equation*}
\end{subequations}
where the product $\Phi_i \left(\zeta_i + \xi_i^s\right)=\Phi_i \xi_i$ gives the vector composed of the stacked water flows at each section of any reach considered within coalition $\mathcal{C}_i$.\par
Next, a detailed description of the proposed algorithm is provided:
\begin{alg} 
\textbf{Coalitional MPC}
\hrule
\begin{enumerate}[Step 1.]
  \item Prepare a set $\mathfrak{L}^{+}$ of suitable network topologies to be evaluated for their use during the next interval $T_{\Lambda}$, defined as a multiple of the sampling time at the bottom layer.
	\item For each $\Lambda \in \mathfrak{L}^{+}$ compute $\mathbb{K}(\Lambda)$ and $\mathbb{P}(\Lambda)$.
	\item For each $\Lambda \in \mathfrak{L}^{+}$, estimate the steady-state effect due to neighbor coalitions as:
	\begin{equation*}
	\hat{\omega}_i(k) = \sum\limits_{j\in \mathcal{N}_{\mathcal{C}_i}}\Xi_{ij}\hat{\xi}^s_j+ \Upsilon_{ij}\hat{\upsilon}^s_j, \; \forall i\in\mathcal{V}_{\Lambda}
	\end{equation*}
	where each pair $\{\hat{\xi}^s_j,\hat{\upsilon}^s_j\}$ is the most updated setpoint available from coalition $\mathcal{C}_j$.
	\item For each $\Lambda \in \mathfrak{L}^{+}$, solve \eqref{eq_setpoint_botlay} to obtain the setpoints $\{\bar{\xi}_i(k),\bar{\upsilon}_i(k)\}$, $\forall i \in \mathcal{V}_{\Lambda}$, using the value of $\hat{\omega}_i(k)$ computed at Step 3.
	\item Compute an estimate of the global cost with \eqref{eq:v_LG} and pick the network mode $\Lambda^{\ast}$ that would give the minimum cost.
	\item Communicate each local feedback law $K_i(\Lambda^{\ast})$ and $P_i(\Lambda^{\ast})$ to their corresponding coalition.
	\item Each coalition computes a setpoint $\{\xi^s_i(k),\upsilon^s_i(k)\}$ by solving \eqref{eq_setpoint_botlay} and \eqref{eq_setpoint_opt}.
	\item Each coalition solves \eqref{eq_decentr_MPC} for $\upsilon'_i(k)$, and the control action \eqref{eq_input_MPC}
	is applied to the subsystems in coalition $\mathcal{C}_i$.
	\item By means of the local Kalman filter, each coalition obtains the estimate $\hat{\omega}_i(k+1) \equiv \hat{\omega}_i(k+1|k)$ to be used during the following time step.
	\item Repeat Steps 7--9 during the interval $T_{\Lambda}$, then go to Step 1.
\end{enumerate}
\hrule
\end{alg}
\section{Case Study: the Dez canal}
\label{sec_canalDez}
In this section we present a case study of the control of water levels in the Dez irrigation canal, to demonstrate the performance of the proposed control scheme.\par
\begin{figure}[tbp]
  \centering
    \includegraphics[width=\columnwidth]{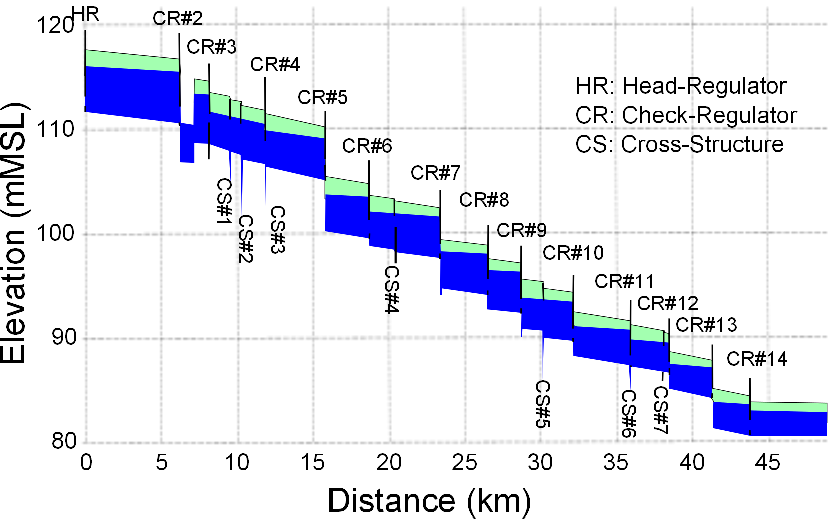}
    \caption{Longitudinal layout of the first 44 km of the west main canal of the Dez irrigation network.}
    \label{fig:dezcanal}
\end{figure}

Located in the south-west of Iran, near the city of Dezful, the Dez canal was designed for the conveyance of irrigation water from a large dam on the Dez river to the irrigated areas in the north of Khuzestan province. For this study we consider a 44 km section of the west main canal, corresponding to 13 pools. Its longitudinal profile is shown in Figure~\ref{fig:dezcanal}.
\subsection{Model: Simulation}
\label{sec_modelsim}
A detailed simulation of the physics of the west main canal of the Dez irrigation network has been performed using the SOBEK modeling suite for water systems. Based on the WL$\vert$Delft Hydraulics implicit finite difference scheme~\cite{manual_sobek}, this software package is currently developed at the Deltares research institute in the Netherlands. The following details about the design of the canal are specified in order to accurately reproduce its dynamics: path, cross-sections, layout of the canal network, type and width of gate, crest levels and discharge coefficients, upstream and downstream boundary conditions. All the necessary information about the geometry of the canal and the hydraulic structures has been obtained from the water authority of Khuzestan province. The boundary condition at the head gate is a constant water level at its upstream side, while its maximum discharge capacity is 157 m$^3$/s. The model has been calibrated and validated using real data relative to six months of operation \cite{Isapoor2011}, and further employed in~\cite{HashemyEtAl2013,HashemyVanOverloop2013}.
\subsection{Model: Control}
\begin{figure}[tbp]
  \centering
    \includegraphics[width=0.7\columnwidth]{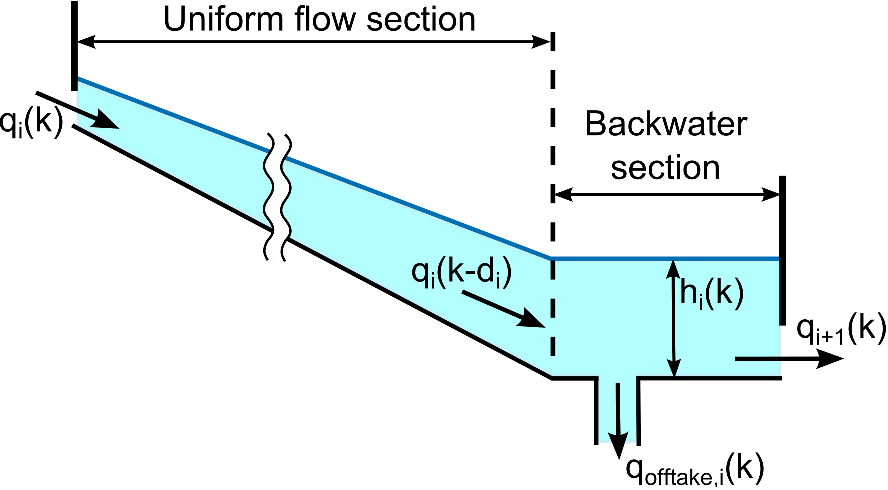}
    \caption{Simplified profile of a reach. The inflow $q_{i}(k)$ crosses the uniform flow section in a time $d_i$. The flow $q_{i}(k-d_i)$ enters the backwater section inducing a change in the water level $h_i(k)$. The water demand is $q_{\mathrm{offtake},i}(k)$, while $q_{i+1}(k)$ is the flow passing to the downstream reach.}
    \label{fig:idmodel}
\end{figure}

A discrete-time linear approximation of the water levels' dynamics, namely the integrator delay (ID) model, is employed in both control layers. Proposed in~\cite{Schuurmans1997}, it is commonly adopted in studies regarding the application of advanced control strategies on water systems (e.g.,~\cite{Negenborn:08c,vanOverloop2010,Clemmens2012,HashemyEtAl2013}) where an essential, minimal order characterization of the response is critical for limiting the computational complexity. According to the ID model, each canal reach is schematized as a uniform flow section followed by a backwater section where the water accumulates maintaining an almost horizontal surface~\cite{Overloop2008a}. Transport delay and storage area are the features parametrized for each reach. The scheme in Figure~\ref{fig:idmodel} illustrates this idea. A change $\Delta q_{i}$ of the inflow is regarded as a kinematic wave traveling along the uniform flow section in downstream direction. The delay $d_i$ is the discretized time interval before $\Delta q_{i}$ induces a variation in the water level $h_i$ in the backwater section. This is considered the actual reservoir of the reach, and constitutes the integrator part of the model, characterized by its average storage area $A_{\mathrm{s},i}$. The offtake flow $q_{\mathrm{offtake},i}(k)$ --- usually scheduled in advance by the authorities in charge of the canal --- is a measurable disturbance.\par
Indeed, the nonlinear dynamics of the system are not covered by this model. When resonance waves play a dominant role --- which is usual in short or flat pools at low discharge rates --- the closed loop system can become unstable. Common ways to deal with this issue are, e.g., low-pass filtering, time-variant linear models, higher order models~\cite{vanOverloop2005,vanOverloop2010}. However, the safety of operation w.r.t. the amplification of resonance waves has been studied beforehand. Notice that the present canal section consists of long and steep pools; also, the two scenarios considered in the remainder reflect high discharge rates.\par
\begin{assum}\label{assum_localflowctr}
The gates are equipped with a local flow controller which manipulates the opening of the gate in order to maintain the water flow at a reference value. If the response of these local control loops is sufficiently fast, the only cause of unintended coupling among adjacent reaches is the manipulation of the flow through a gate, that will affect the water level in the upstream reach \cite{Schuurmans1997}.
\end{assum}
The most common technique used in primary irrigation canals, and considered in this case study as well, is the distant downstream control. According to this technique, the water level in the backwater section of a reach is controlled by manipulating the opening of the upstream gate, physically located at the end of the upstream reach. For example, in response to a decrease of the water level, the gate upstream will open in order to restore it. From Assumption~\ref{assum_localflowctr} it follows that this will produce a \emph{direct} unintended effect on the water level upstream.\par
For every reach, the following variables are considered:
\begin{itemize}
	\item $e_i \equiv h_i - \bar{h}_i$, the error with respect to the desired water level, in the backwater section of reach $i$;
	\item $q_i$, the input flow to reach $i$.
\end{itemize}
In order to take into account the delay $d_i$ along the uniform flow section, an augmented state representation is used. Based on the ID model, the following discrete linear time-invariant model is obtained:
\begin{equation}
\label{eq:idmodelfor}
\begin{split}
e_i(k+1) & = e_i(k) + \frac{T_{\mathrm{c}}}{A_{\mathrm{s},i}} \big[q_{i}(k-d_i) - q_{\mathrm{out},i}(k)\big]\\
q_i(k) & = q_i(k-1) + \Delta q_i(k)
\end{split}
\end{equation}
with
\[
q_{\mathrm{out},i}(k) = q_{i+1}(k-1) + \Delta q_{i+1}(k) + q_{\mathrm{offtake},i}(k)
\]
where $\Delta q_i \equiv u_i$ is a component of the control action computed through~\eqref{eq_input_MPC}, representing the increment in the target flow of the gate's local controller, $A_{\mathrm{s},i}$ is the backwater surface area, $d_i$ is the discretized value of the transport delay and $T_{\mathrm{c}}$ is the sampling time.\par
The water levels are subject to an offset caused by the offtake flows and the disturbance due to downstream reaches. The purpose of the proposed control scheme is to maintain the water level of each reach around a fixed value ($\bar{h}_i$), that is, to regulate the level errors $e_i$ to zero, while minimizing the control effort and the number of active network links. This task is accomplished by driving the system's state to the setpoint computed through \eqref{eq_setpoint_botlay} and \eqref{eq_setpoint_opt}.\par
The parameters of the ID model for the 13 reaches have been identified through simulations on the validated model in SOBEK (see \S\ref{sec_modelsim}), and are given in Table~\ref{tab_param}. The transport delay and the average storage area have been characterized individually for each reach, considering the canal at steady state with an input amounting to 80\% of the maximum inflow, and all the offtakes at 80\% of the maximum discharge capacity. The values in Table~\ref{tab_param} refer to the canal in this operational conditions. An additional identification of the model parameters has been carried out for a medium discharge setting, namely 50\% of the maximum inflow and 50\% of the maximum offtake. Nevertheless, the parameters have demonstrated little sensitivity to the change in the discharge regime. Thus, neglecting the dependence of the parameters on the flow, the values in Table~\ref{tab_param} have been used through the scenarios presented in \S~\ref{sec_sim}. Notice that the implementation of an offset-free method, together with the inherent robustness of the MPC, contributes to the compensation of model-plant mismatches.
\begin{table}[tbp]
\centering
   \begin{tabular}{c >{\centering}m{7.5ex} >{\centering}m{11ex} >{\centering}m{11ex} m{6ex}<{\centering} }
     Reach & Length [m] & Width (bottom) [m] & Backwater surface ($\cdot 10^5$) [m$^2$]& Delay steps \\ \hline
     1 & 6219 & 12 & 0.9318 & 3 \\ 
     2 & 1933 & 12 & 1.0952 & 1 \\ 
     3 & 3718 & 10 & 0.8554 & 2 \\ 
     4 & 3906 & 10 & 3.7060 & 2 \\ 
     5 & 2934 & 5 & 1.7095 & 2 \\ 
     6 & 4670 & 5 & 0.7786 & 3 \\ 
     7 & 3110 & 5 & 0.6661 & 2 \\
     8 & 2240 & 5 & 0.8904 & 1 \\ 
     9 & 3405 & 5 & 0.8671 & 2 \\ 
     10 & 3820 & 5 & 0.4897 & 2 \\ 
     11 & 2520 & 4 & 0.4032 & 2 \\
     12 & 2874 & 4 & 0.3820 & 2 \\
     13 & 2468 & 5 & 0.3884 & 2 \\
     \hline
   \end{tabular}
   \caption{Parameters of the first 13 reaches of the west main Dez canal, identified at 80\% of the maximum discharge rate.}
	\label{tab_param}
\end{table}%
\subsection{Simulation}
\label{sec_sim}
\begin{figure}[tbp]
  \centering
   \includegraphics[width=0.8\columnwidth]{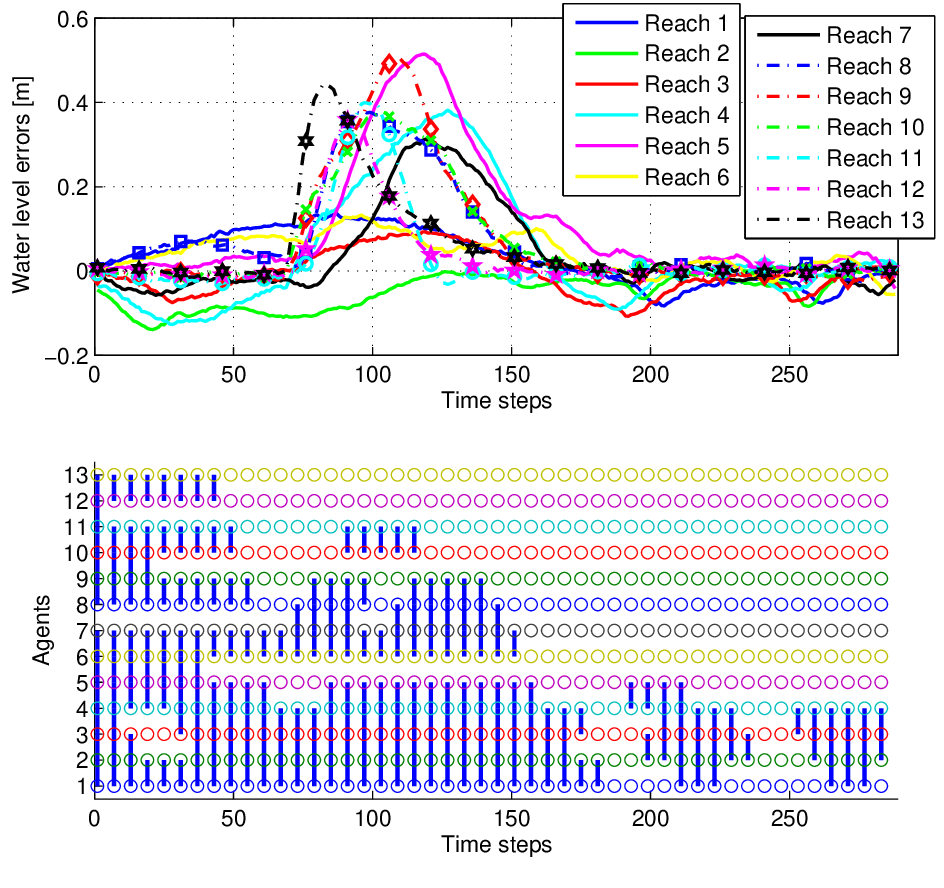}
   \caption{Scenario 1: At $k = 72$ reaches 4, 9, 10 and 13 undergo a step decrease in the offtake flows. The upper plot shows how the controller regulates the water level errors along the canal. Each blue line in the bottom plot represents an active data link between two control agents. Starting from a centralized configuration, links are deactivated one at a time until the variation in the offtakes is sensed. Then links are enabled to form coalitions among the most concerned control agents, until the disturbance is eventually rejected.}
    \label{fig:scen1}
\end{figure}

Two scenarios are analyzed in the remainder, reflecting the canal operation at medium--high discharge regimes. In the first one, the water levels and flows are settled, supplying some constant nonzero offtakes along the canal until, 360 minutes past the beginning of the simulation ($k=72$), four of the reaches undergo a step decrease in their offtake flows. The same situation is also considered in the second scenario, with the addition of a second step change --- 360 minutes after the first one ($k=144$) --- that restores the offtakes to their former values. The variation of the offtake flows amounts to 10 m$^3$/s at reaches 4 and 13, and to 5 m$^3$/s at reaches 9 and 10, ranging from 20\% to 100\% of their initial magnitude. The simulation is meant to test the performance of the proposed control scheme in rejecting the simultaneous changes in the offtakes, keeping the water level at each reach around its corresponding reference. Notice that the exchange of information between different agents is allowed or forbidden according to the network topology chosen, on the basis of how the dynamic coupling between the reaches evolve with time. This operation is carried out at the top layer with a coarser sampling time w.r.t.~the one used at the bottom layer (see \S \ref{sec_toplayer}). In the simulations, the state of only one of the links is allowed to change between any two subsequent choices of topology. Hard constraints are imposed on the water flow increment at each gate, $\left|\Delta q_i(k)\right| \leq 1$ m$^3$/s. Constraints on the direction of the water flows, i.e., $q_i(k)>0$, are imposed as soft constraints. Table \ref{tab_contr_par} lists the values of the controller's parameters used for both scenarios. These values have been tuned by trial and error, balancing the performance improvement with the computational requirements.\footnote{An initial guess for the values of matrices $Q_i$ and $R_i$ has been set according to the inverse-variance weighting method.}
\begin{table}[tbp]
\centering
   \begin{tabular}{c c c}
     Symbol & Description & Value \\ \hline
     $N_p$ & Prediction horizon & 10 \\ 
     $N_c$ & Control horizon & 3 \\
     $Q_i$ & Weight on the errors & 250 \\ 
     $R_i$ & Weight on the flow increments & 2800 \\ 
     $S_i$ & Weight on flows $<0$ (soft constr.) & $10^4$ \\ 
     $c_{\ell}$ & Cost of an active link & $0.6$ \\
     $T_{\mathrm{c}}$ & Sampling time [s] & $300$ \\
     \hline
   \end{tabular}
   \caption{Parameters of the controller.}
	\label{tab_contr_par}
\end{table}
\begin{rem}
In practice, the agents are provided in advance with the future offtakes' schedule. In this test, however, only offtakes measured in place at each sample time are considered. This allows us to show the effectiveness of the proposed control scheme in an ``on demand'' operation, in which users can take water anytime without any previous agreement with canal authorities.
\end{rem}
The plots in Figure~\ref{fig:scen1} refer to Scenario 1. In the upper plot, the evolution of water level errors is shown. It can be seen how the sudden decrease in the offtake flows at time $k=72$ causes the water levels to rise above the desired setpoint, reaching in some pools a peak of 0.5~m. In the bottom plot of Figure~\ref{fig:scen1}, blue lines indicate active links. The agents in charge of the first four upstream reaches act jointly when the disturbance occurs. This coordination allow to minimize the perturbation on reach 3 (and consequently on reach 2 and 1) due to the recovery maneuver of the agent in charge of reach 4.\par
\begin{figure}[tbp]
  \centering
   \includegraphics[trim = 0 0 0.5cm 0, width=0.8\columnwidth]{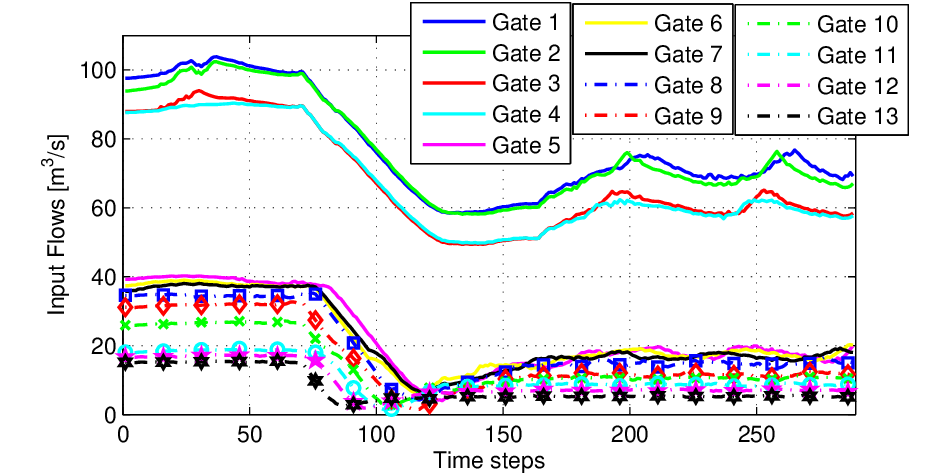}
   \caption{Scenario 1: Water flows in input to each reach.}
   \label{fig:scen1_flow}
\end{figure}

During the transient response of the local flow controllers, the sudden variation of the offtake flows produces an immediate perturbation in downstream direction. Therefore, in order to improve the performance of the overall system, the agents are organized into bigger coalitions while the effect of the offtakes' variation is perceived. As the disturbance is rejected, the data links are disabled and most of the agents continue to control their reaches in a decentralized way. It can be seen that a coalition among upstream agents is profitable and thus does not eventually disappear. One reason to this might be the fact that, in the system under study, an accurate decentralized estimation of neighbor's state is not possible, and water levels at upstream reaches tend to deviate from the setpoint as soon as their agents stop to communicate. This may suggest that in an optimal partition of the overall system the upstream agents would be bound into the same coalition. Figure \ref{fig:scen1_flow} shows the inflows $q_i(k)$ to the reaches. Starting from time $k=72$, the flows are reduced in response to the decreased offtakes in order to bring the water levels back to their setpoints.\par
\begin{figure}[tbp]
  \centering
   \includegraphics[width=0.8\columnwidth]{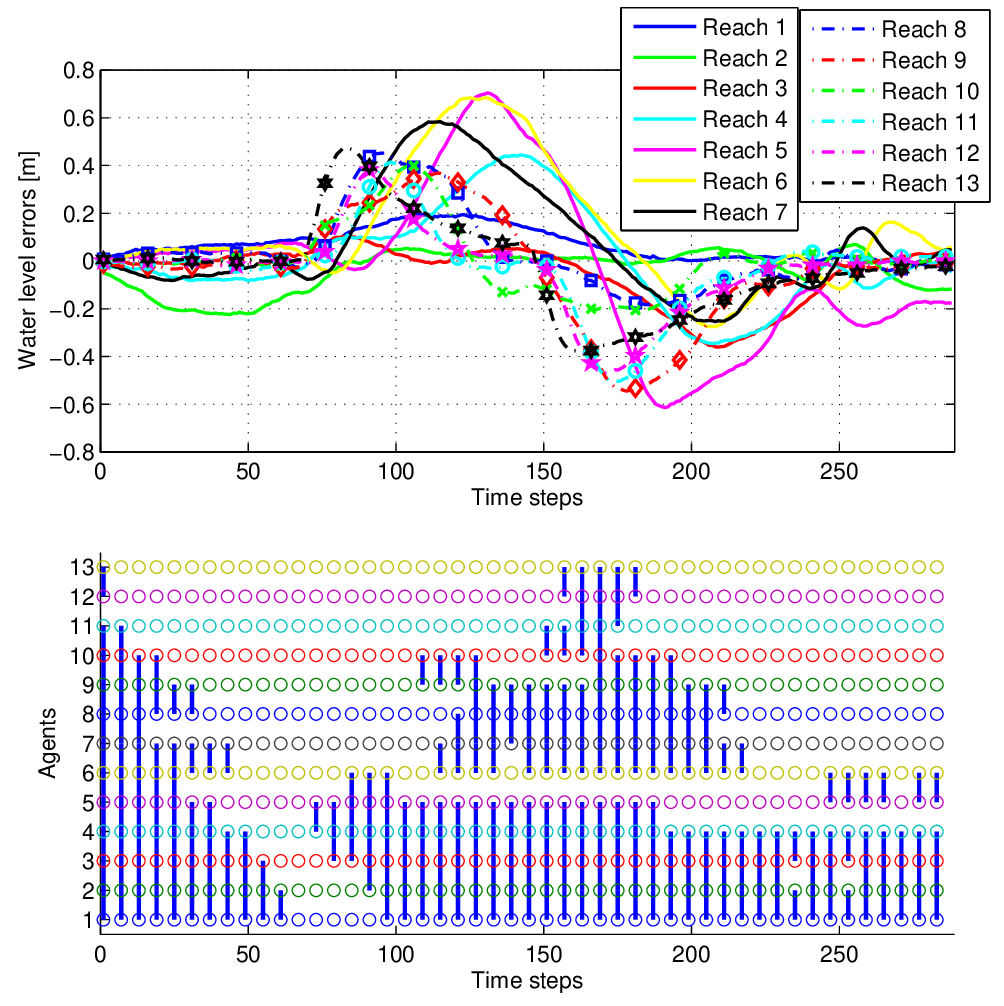}
   \caption{Scenario 2: The offtake flows in reaches 4, 9, 10 and 13 undergo a step decrease at $k = 72$, and are restored at $k = 144$. The upper plot shows the water level errors in all the reaches. The use of data links between the control agents is represented by the blue lines in the bottom plot. With the water levels at their setpoints, the network changes toward a decentralized topology. In reaction to the offtake change, the control agents are organized into coalitions to improve the overall response.}
    \label{fig:scen2}
\end{figure}
\begin{figure}[tbp]
  \centering
   \includegraphics[trim = 0 0 0.5cm 0, width=0.8\columnwidth]{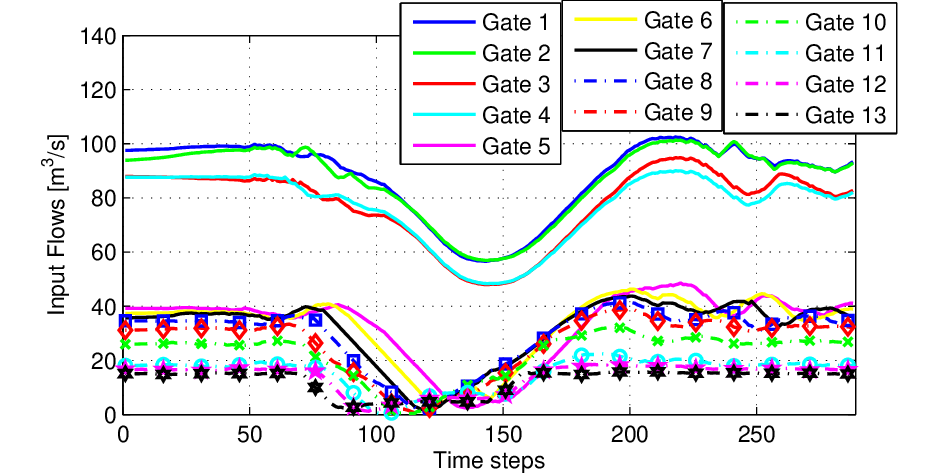}
   \caption{Scenario 2: Water flows in input to each reach.}
   \label{fig:scen2_flow}
\end{figure}

The results relative to Scenario 2 are shown in Figure \ref{fig:scen2}. Before the offtakes variation at time $k = 72$, the network topology changes to a decentralized configuration while the water levels are kept at their setpoints. When the decrease in the offtakes occurs, the system reacts by reducing the inflows (Fig.~\ref{fig:scen2_flow}), while data links are enabled in order to coordinate the operations along the canal. At time $k = 144$ the offtake flows are restored, which is matched with an increase in the input flows. As the error in the water levels is attenuated, downstream data links are disabled. Notice in the last part of the simulation the persistence of the coalition among the upstream agents, and also the coalition formed by agents 5 and 6, since the water level in reach 5 does not converge to its setpoint.\par
It can be seen in Figures \ref{fig:scen1_flow} and \ref{fig:scen2_flow} that the constraints on the water flows are satisfied in both scenarios.\par
\begin{figure}[tbp]
  \centering
   \includegraphics[trim = 0 0 0.5cm 0, width=0.8\columnwidth]{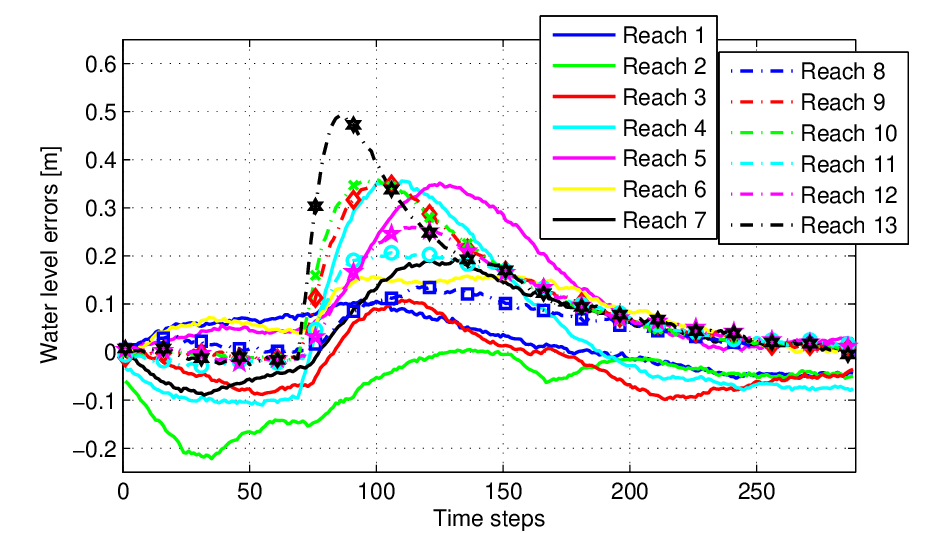}
   \caption{Scenario 1: Water level errors with centralized MPC controller.}
   \label{fig:scen1_centr}
\end{figure}
\begin{figure}[tbp]
  \centering
   \includegraphics[trim = 0 0 0.5cm 0, width=0.8\columnwidth]{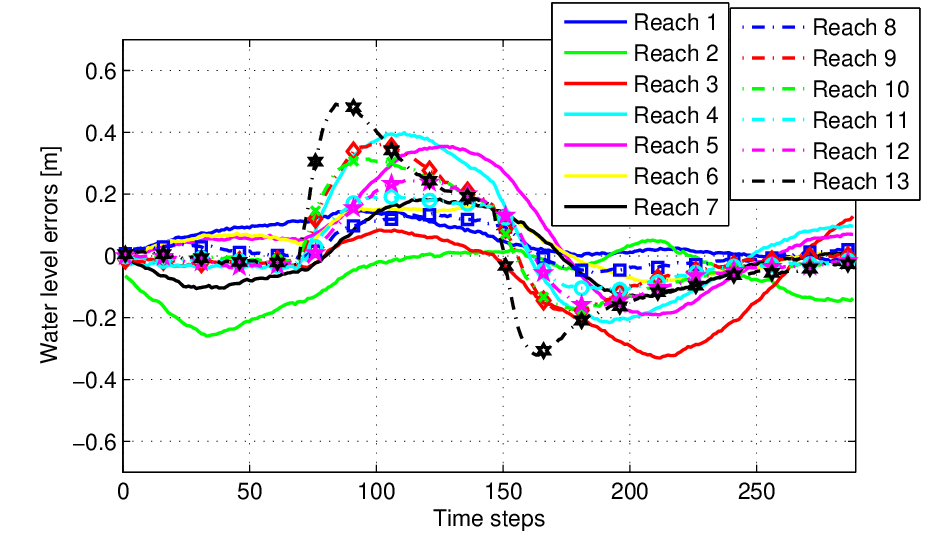}
   \caption{Scenario 2: Water level errors with centralized MPC controller.}
   \label{fig:scen2_centr}
\end{figure}
\begin{figure}[tbp]
  \centering
   \includegraphics[trim = 0 0 0.5cm 0, width=0.8\columnwidth]{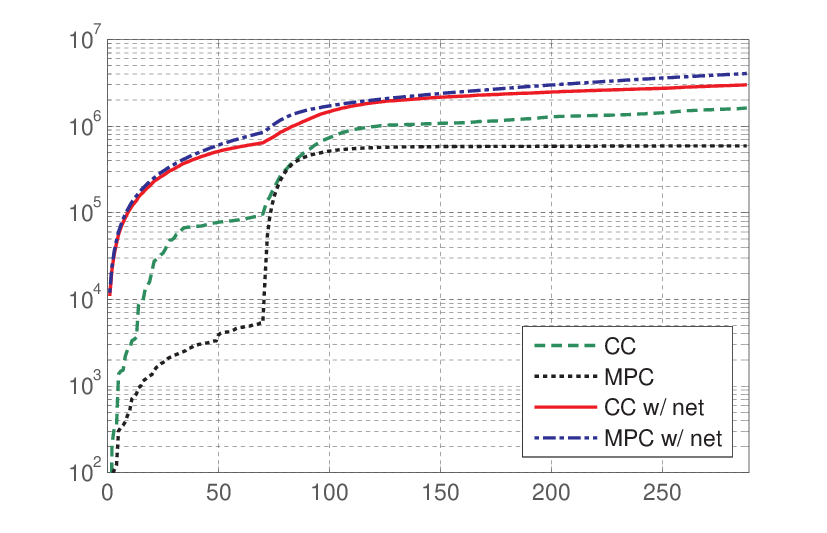}
   \caption{Scenario 1: Accumulated costs comparison between the proposed coalitional control strategy and a centralized MPC, w/ and w/o network costs.}
   \label{fig:sc1_comp_costes}
\end{figure}

For comparison, the performance of a centralized MPC controller in the same scenarios is shown in Figures \ref{fig:scen1_centr} and \ref{fig:scen2_centr}. The centralized control law is computed as in \eqref{eq_input_MPC}, using the same parameters of Table \ref{tab_contr_par} (except for the cost of active links, which is set to zero). The centralized controller can coordinate the response of the entire canal to provide a faster reaction to the disturbance, which yields about 30\% reduction in the level error peaks.\footnote{An exception to this is the water level in reach 13, as no further downstream reaches can be employed to improve its response (the water discharge at the downstream gate in reach 13 cannot be manipulated by the controller).} As expected, due to the absence of network topology switching with centralized control, a smoother response of the system is obtained. Notice that, even without taking into account any offtake forecast, the results achieved with the proposed distributed control scheme are within the admitted range of canal operation. Table~\ref{tab_costes} displays a comparison of the average performance costs of the proposed coalitional scheme and the centralized MPC, for both scenarios. In Figure~\ref{fig:sc1_comp_costes} the accumulated costs relative to Scenario 1 are represented, evidencing the impact of communication-related costs in a centralized framework. The average number of decision variables for the decentralized MPC problems solved by the coalitions at the bottom layer is shown in Table~\ref{tab_decvar}, along with the average number of coalitions.
\begin{table}[tbp]
\centering
   \begin{tabular}{c l r@{.}l r@{.}l}
      & & Coal &~$( \cdot 10^3 )$ & Centr &~$( \cdot 10^3 )$ \\ \hline
     \multirow{2}{*}{Scn. 1} & $c_{\ell} = 0$ & 5&62 & 2&06 \\ 
     & $c_{\ell} = 0.6$ & 10&44 & 14&06 \\ \hline
     \multirow{2}{*}{Scn. 2} & $c_{\ell} = 0$ & 9&36 & 3&89  \\ 
     & $c_{\ell} = 0.6$ & 14&80 & 15&89 \\ 
     \hline
   \end{tabular}
   \caption{Comparison of the average costs of the proposed coalitional scheme and centralized MPC for the two scenarios.}
	\label{tab_costes}
\end{table}
\begin{table}[tbp]
\centering
   \begin{tabular}{c c c}
      & $n^{\circ}$ dec. var. & $n^{\circ}$ coal. \\ \hline
     Scn. 1 & 5.6 & 8.2 \\ 
     Scn. 2 & 6.3 & 7.6  \\ \hline
     Centr. MPC & 39 & 1 \\ 
     \hline
   \end{tabular}
   \caption{Average number of decision variables for the decentralized MPC controllers at the bottom layer, relative to both scenarios.}
	\label{tab_decvar}
\end{table}
\section{Conclusions}
We have presented a hierarchical control scheme aimed to large-scale networked systems, that dynamically adapts the information flow between agents as a function of the state of the system and the cost of use of the network resources. The proposed scheme promotes \emph{cooperation} between different agents whenever it results in a sensible performance improvement. The controller have been tested within a case study of the control of water levels in the Dez irrigation canal, using the accurate hydrodynamic simulation environment SOBEK.\par
Although suboptimal w.r.t.~a centralized MPC control, the performance of the proposed strategy lies within the admitted range of operation of the canal. On the other hand, the need of constant exchange of information through the entire system is obviated. Also, a sensible reduction of the computational requirements can be attained in high-dimensional systems. Regarding the network topology optimization, \emph{(i)} the evaluation of different topologies is independent and can be executed in parallel on a multi-processor platform, and \emph{(ii)} the control laws resulting from the evaluation of any network topology could be stored and reused. Another advantage of the proposed control strategy over a centralized controller is the possibility to explicitly manage any link failure by switching to a different network configuration, improving the robustness to network-related issues.\par
\begin{figure}[tbp]
  \centering
    \includegraphics[width=0.9\columnwidth]{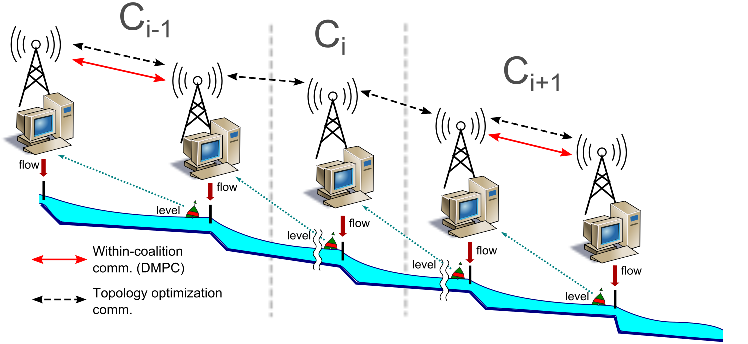}
    \caption{Scheme of a possible implementation of the proposed control strategy.}
    \label{fig_impl}
\end{figure}

Given the availability of a communication infrastructure (e.g., point-to-point wireless network) across the canal, both layers of the proposed scheme can be distributed among the local control agents (Fig.~\ref{fig_impl}). The network topology optimization task can be shared, e.g., performing the distributed synthesis of the global feedback law with the method presented in \cite{ConteEtAl2012_ACC, ConteEtAl2013_ECC}. Alternatively, one of the agents can assume the role of supervisor, and carry out the topology optimization as well as its bottom layer control tasks. Both solutions will require data exchange among all the agents before each topology switch. Within members of a given coalition, a higher rate of communication is required for the decentralized MPC at the bottom layer.\par
Future work shall focus on how the switching network topology influence the global state estimation. In the proposed strategy, deviations of the water levels w.r.t. to their initial values have been penalized through the cost functions. Constraints on the water level have not been considered to avoid feasibility issues (mainly caused by the drifts in the state estimate occurring at each topology switch). Moreover, due to the variance in the estimate, a robust design of the controller and the observer is in order for the correct handling of state constraints. For these same reasons, soft constraints have been chosen for the water flows.
Further investigation shall also address the coalition formation process --- including the stability of any given configuration --- and its relation with the optimal system decomposition.
%
\newpage
\small
\bibliographystyle{model1-num-names}
\bibliography{bibpepe}

\begin{thebibliography}{41}
\expandafter\ifx\csname natexlab\endcsname\relax\def\natexlab#1{#1}\fi
\providecommand{\bibinfo}[2]{#2}
\ifx\xfnm\relax \def\xfnm[#1]{\unskip,\space#1}\fi
\bibitem[{Fele et~al.(2014)Fele, Maestre, Hashemy, {Mu\~{n}oz de la Pe\~{n}a},
  and Camacho}]{Fele_2014JPC}
\bibinfo{author}{F.~Fele}, \bibinfo{author}{J.~M. Maestre},
  \bibinfo{author}{S.~M. Hashemy}, \bibinfo{author}{D.~{Mu\~{n}oz de la
  Pe\~{n}a}}, \bibinfo{author}{E.~F. Camacho},
\newblock \bibinfo{title}{Coalitional model predictive control of an irrigation
  canal},
\newblock \bibinfo{journal}{Journal of Process Control} \bibinfo{volume}{24}
  (\bibinfo{year}{2014}) \bibinfo{pages}{314 -- 325}.
\bibitem[{Negenborn et~al.(2010)Negenborn, Lukszo, and
  Hellendoorn}]{NEGENBORN10BOOK}
\bibinfo{editor}{R.~R. Negenborn}, \bibinfo{editor}{Z.~Lukszo},
  \bibinfo{editor}{H.~Hellendoorn} (Eds.), \bibinfo{title}{Intelligent
  Infrastructures}, \bibinfo{publisher}{Springer}, \bibinfo{year}{2010}.
\bibitem[{Malaterre et~al.(1998)Malaterre, Rogers, and
  Schuurmans}]{MALATERRE98JIDE}
\bibinfo{author}{P.~Malaterre}, \bibinfo{author}{D.~Rogers},
  \bibinfo{author}{J.~Schuurmans},
\newblock \bibinfo{title}{Classification of canal control algorithms},
\newblock \bibinfo{journal}{Journal of Irrigation and Drainage Engineering}
  \bibinfo{volume}{124} (\bibinfo{year}{1998}) \bibinfo{pages}{3--10}.
\bibitem[{Clemmens and Schuurmans(2004)}]{CLE04JIDE}
\bibinfo{author}{A.~J. Clemmens}, \bibinfo{author}{J.~Schuurmans},
\newblock \bibinfo{title}{Simple optimal downstream feedback canal controllers:
  Theory},
\newblock \bibinfo{journal}{Journal of Irrigation and Drainage Engineering}
  \bibinfo{volume}{130} (\bibinfo{year}{2004}) \bibinfo{pages}{26--34}.
\bibitem[{van Overloop(2006)}]{PJVO06TH}
\bibinfo{author}{P.~J. van Overloop}, \bibinfo{title}{Model Predictive Control
  on Open Water Systems}, Ph.D. thesis, Delft University of Technology,
  \bibinfo{address}{Delft, The Netherlands}, \bibinfo{year}{2006}.
\bibitem[{Scattolini(2009)}]{SCA09JPC}
\bibinfo{author}{R.~Scattolini},
\newblock \bibinfo{title}{Architectures for distributed and hierarchical model
  predictive control - a review},
\newblock \bibinfo{journal}{Journal of Process Control} \bibinfo{volume}{19}
  (\bibinfo{year}{2009}) \bibinfo{pages}{723--731}.
\bibitem[{Negenborn et~al.(2009)Negenborn, {van Overloop}, Keviczky, and {De
  Schutter}}]{Negenborn:08c}
\bibinfo{author}{R.~R. Negenborn}, \bibinfo{author}{P.~J. {van Overloop}},
  \bibinfo{author}{T.~Keviczky}, \bibinfo{author}{B.~{De Schutter}},
\newblock \bibinfo{title}{Distributed model predictive control for irrigation
  canals},
\newblock \bibinfo{journal}{Networks and Heterogeneous Media}
  \bibinfo{volume}{4} (\bibinfo{year}{2009}) \bibinfo{pages}{359--380}.
\bibitem[{Fawal et~al.(1998)Fawal, Georges, and Bornard}]{Fawal1998}
\bibinfo{author}{H.~Fawal}, \bibinfo{author}{D.~Georges},
  \bibinfo{author}{G.~Bornard},
\newblock \bibinfo{title}{Optimal control of complex irrigation systems via
  decomposition-coordination and the use of augmented lagrangian},
\newblock in: \bibinfo{booktitle}{1998 IEEE International Conference on
  Systems, Man, and Cybernetics}, volume~\bibinfo{volume}{4}, pp.
  \bibinfo{pages}{3874--3879}.
\bibitem[{Zafra~Cabeza et~al.(2011)Zafra~Cabeza, Maestre, Ridao, Camacho, and
  S\'anchez}]{ZAFRAJMM10JPC}
\bibinfo{author}{A.~Zafra~Cabeza}, \bibinfo{author}{J.~M. Maestre},
  \bibinfo{author}{M.~A. Ridao}, \bibinfo{author}{E.~F. Camacho},
  \bibinfo{author}{L.~S\'anchez},
\newblock \bibinfo{title}{A hierarchical distributed model predictive control
  approach in irrigation canals: A risk mitigation perspective},
\newblock \bibinfo{journal}{Journal of Process Control} \bibinfo{volume}{21}
  (\bibinfo{year}{2011}) \bibinfo{pages}{787--799}.
\bibitem[{\v{S}iljak(1991)}]{Siljak1991}
\bibinfo{author}{D.~D. \v{S}iljak}, \bibinfo{title}{Decentralized control of
  complex systems}, \bibinfo{publisher}{Boston: Academic Press Inc.},
  \bibinfo{year}{1991}.
\bibitem[{Negenborn et~al.(2006)Negenborn, De~Schutter, and Hellendoorn}]{NEG1}
\bibinfo{author}{R.~R. Negenborn}, \bibinfo{author}{B.~De~Schutter},
  \bibinfo{author}{H.~Hellendoorn},
\newblock \bibinfo{title}{Multi-agent model predictive control of
  transportation networks},
\newblock in: \bibinfo{booktitle}{Proceedings of the 2006 IEEE International
  Conference on Networking, Sensing and Control (ICNSC 2006)},
  \bibinfo{address}{Ft. Lauderdale, Florida}, pp. \bibinfo{pages}{pp.
  296--301}.
\bibitem[{Venkat(2006)}]{VEN06TH}
\bibinfo{author}{A.~N. Venkat}, \bibinfo{title}{Distributed Model Predictive
  Control: Theory and Applications}, Ph.D. thesis, University of
  Wisconsin-Madison, \bibinfo{year}{2006}.
\bibitem[{Maestre et~al.(2011)Maestre, Mu\~noz de~la Pe\~{n}a, and
  Camacho}]{JMM10OCAM}
\bibinfo{author}{J.~M. Maestre}, \bibinfo{author}{D.~Mu\~noz de~la Pe\~{n}a},
  \bibinfo{author}{E.~F. Camacho},
\newblock \bibinfo{title}{Distributed model predictive control based on a
  cooperative game},
\newblock \bibinfo{journal}{Optimal Control Applications and Methods}
  \bibinfo{volume}{32} (\bibinfo{year}{2011}) \bibinfo{pages}{153--176}.
\bibitem[{Farina and Scattolini(2012)}]{Farina:2012:DPC}
\bibinfo{author}{M.~Farina}, \bibinfo{author}{R.~Scattolini},
\newblock \bibinfo{title}{Distributed predictive control: A non-cooperative
  algorithm with neighbor-to-neighbor communication for linear systems},
\newblock \bibinfo{journal}{Automatica} \bibinfo{volume}{48}
  (\bibinfo{year}{2012}) \bibinfo{pages}{1088--1096}.
\bibitem[{Alessio and Bemporad(2007)}]{ALE07ECC}
\bibinfo{author}{A.~Alessio}, \bibinfo{author}{A.~Bemporad},
\newblock \bibinfo{title}{Decentralized model predictive control of constrained
  linear systems},
\newblock in: \bibinfo{booktitle}{Proceedings of the 2007 European Control
  Conference}, pp. \bibinfo{pages}{2813--2818}.
\bibitem[{Magni and Scattolini(2006)}]{MS06Automatica}
\bibinfo{author}{L.~Magni}, \bibinfo{author}{R.~Scattolini},
\newblock \bibinfo{title}{Stabilizing decentralized model predictive control of
  nonlinear systems},
\newblock \bibinfo{journal}{Automatica} \bibinfo{volume}{42}
  (\bibinfo{year}{2006}) \bibinfo{pages}{1231--1236}.
\bibitem[{Trodden and Richards(2009)}]{Trodden09}
\bibinfo{author}{P.~Trodden}, \bibinfo{author}{A.~Richards},
\newblock \bibinfo{title}{Adaptive cooperation in robust distributed model
  predictive control},
\newblock in: \bibinfo{booktitle}{Control Applications, (CCA) Intelligent
  Control, (ISIC), 2009 IEEE}, pp. \bibinfo{pages}{896--901}.
\bibitem[{Jilg and Stursberg(2013)}]{JilgStur2013_IFAC}
\bibinfo{author}{M.~Jilg}, \bibinfo{author}{O.~Stursberg},
\newblock \bibinfo{title}{{Hierarchical Distributed Control for Interconnected
  Systems}},
\newblock in: \bibinfo{booktitle}{13th IFAC Symposium on Large Scale Complex
  Systems: Theory and Applications}, volume~\bibinfo{volume}{13},
  \bibinfo{address}{Shanghai, China}, pp. \bibinfo{pages}{419--425}.
\bibitem[{Ocampo-Martinez et~al.(2012)Ocampo-Martinez, Barcelli, Puig, and
  Bemporad}]{Ocampo2012HierarcMPC}
\bibinfo{author}{C.~Ocampo-Martinez}, \bibinfo{author}{D.~Barcelli},
  \bibinfo{author}{V.~Puig}, \bibinfo{author}{A.~Bemporad},
\newblock \bibinfo{title}{Hierarchical and decentralised model predictive
  control of drinking water networks: Application to barcelona case study},
\newblock \bibinfo{journal}{Control Theory Applications, IET}
  \bibinfo{volume}{6} (\bibinfo{year}{2012}) \bibinfo{pages}{62--71}.
\bibitem[{N\'u\~nez et~al.(2013)N\'u\~nez, Ocampo-Mart\'{i}nez, De~Schutter,
  Valencia, L\'{o}pez, and Espinosa}]{NunezEtAl2013}
\bibinfo{author}{A.~N\'u\~nez}, \bibinfo{author}{C.~Ocampo-Mart\'{i}nez},
  \bibinfo{author}{B.~De~Schutter}, \bibinfo{author}{F.~Valencia},
  \bibinfo{author}{J.~L\'{o}pez}, \bibinfo{author}{J.~Espinosa},
\newblock \bibinfo{title}{{A multiobjective-based switching topology for
  hierarchical model predictive control applied to a hydro-power valley}},
\newblock in: \bibinfo{booktitle}{3rd IFAC International Conference on
  Intelligent Control and Automation Science}, \bibinfo{address}{Chengdu,
  China}, pp. \bibinfo{pages}{529--534}.
\bibitem[{Maestre et~al.(2013)Maestre, Mu\~noz de~la Pe\~na, Jim\'enez~Losada,
  Algaba, and Camacho}]{Maestre2013OCAM}
\bibinfo{author}{J.~M. Maestre}, \bibinfo{author}{D.~Mu\~noz de~la Pe\~na},
  \bibinfo{author}{A.~Jim\'enez~Losada}, \bibinfo{author}{E.~Algaba},
  \bibinfo{author}{E.~F. Camacho},
\newblock \bibinfo{title}{A coalitional control scheme with applications to
  cooperative game theory},
\newblock \bibinfo{journal}{Optimal Control Applications and Methods}
  (\bibinfo{year}{2013}).
\bibitem[{Saad et~al.(2009)Saad, Han, Debbah, Hjorungnes, and
  Ba\c{s}ar}]{SAAD09IEEESPM}
\bibinfo{author}{W.~Saad}, \bibinfo{author}{Z.~Han},
  \bibinfo{author}{M.~Debbah}, \bibinfo{author}{A.~Hjorungnes},
  \bibinfo{author}{T.~Ba\c{s}ar},
\newblock \bibinfo{title}{Coalitional game theory for communication networks},
\newblock \bibinfo{journal}{IEEE Signal Processing Magazine, Special Issue on
  Game Theory} \bibinfo{volume}{26} (\bibinfo{year}{2009})
  \bibinfo{pages}{77--97}.
\bibitem[{Fele et~al.(2013)Fele, Maestre, Muros, and Camacho}]{FELE13ICSNC}
\bibinfo{author}{F.~Fele}, \bibinfo{author}{J.~M. Maestre},
  \bibinfo{author}{F.~J. Muros}, \bibinfo{author}{E.~F. Camacho},
\newblock \bibinfo{title}{Coalitional control: an irrigation canal case study},
\newblock in: \bibinfo{booktitle}{Proceedings of the 10th IEEE International
  Conference on Networking, Sensing and Control}, \bibinfo{address}{Evry,
  France}, pp. \bibinfo{pages}{759 -- 764}.
\bibitem[{Camacho and Bordons(2004)}]{EFC1}
\bibinfo{author}{E.~F. Camacho}, \bibinfo{author}{C.~Bordons},
  \bibinfo{title}{Model Predictive Control in the Process Industry. Second
  Edition.}, \bibinfo{publisher}{Springer-Verlag}, \bibinfo{address}{London,
  England}, \bibinfo{year}{2004}.
\bibitem[{Rawlings and Mayne(2009)}]{RawlingsLIB09}
\bibinfo{author}{J.~B. Rawlings}, \bibinfo{author}{D.~Q. Mayne},
  \bibinfo{title}{Model Predictive Control: Theory and Design},
  \bibinfo{publisher}{Nob Hill Publishing}, \bibinfo{year}{2009}.
\bibitem[{Schuurmans(1997)}]{Schuurmans1997}
\bibinfo{author}{J.~Schuurmans}, \bibinfo{title}{Control of Water Levels in
  Open-Channels}, Ph.D. thesis, TUDelft, \bibinfo{year}{1997}.
\bibitem[{Slikker and van~den Nouweland(2001)}]{MSAVDN01}
\bibinfo{author}{M.~Slikker}, \bibinfo{author}{A.~van~den Nouweland},
  \bibinfo{title}{Social and Economics Networks in Cooperative Game Theory},
  \bibinfo{publisher}{Kluwer Academic Publishers}, \bibinfo{year}{2001}.
\bibitem[{Rahwan et~al.(2012)Rahwan, Michalak, Wooldridge, and
  Jennings}]{Rahwan2012}
\bibinfo{author}{T.~Rahwan}, \bibinfo{author}{T.~Michalak},
  \bibinfo{author}{M.~Wooldridge}, \bibinfo{author}{N.~R. Jennings},
\newblock \bibinfo{title}{Anytime coalition structure generation in multi-agent
  systems with positive or negative externalities},
\newblock \bibinfo{journal}{Artificial Intelligence} \bibinfo{volume}{186}
  (\bibinfo{year}{2012}) \bibinfo{pages}{95 -- 122}.
\bibitem[{Maestre et~al.(2011)Maestre, Mu\~noz de~la Pe\~na, Jim\'enez~Losada,
  Algaba, and Camacho}]{MAESTRE11IFAC}
\bibinfo{author}{J.~M. Maestre}, \bibinfo{author}{D.~Mu\~noz de~la Pe\~na},
  \bibinfo{author}{A.~Jim\'enez~Losada}, \bibinfo{author}{E.~Algaba},
  \bibinfo{author}{E.~F. Camacho},
\newblock \bibinfo{title}{An application of cooperative game theory to
  distributed control},
\newblock in: \bibinfo{booktitle}{Proceedings of the 18th IFAC World Congress},
  \bibinfo{address}{Milano, Italy}, pp. \bibinfo{pages}{9121--9126}.
\bibitem[{Kothare et~al.(1996)Kothare, Balakrishnan, and Morari}]{KBM96}
\bibinfo{author}{M.~V. Kothare}, \bibinfo{author}{V.~Balakrishnan},
  \bibinfo{author}{M.~Morari},
\newblock \bibinfo{title}{Robust constrained model predictive control using
  linear matrix inequalities},
\newblock \bibinfo{journal}{Automatica} \bibinfo{volume}{32}
  (\bibinfo{year}{1996}) \bibinfo{pages}{1361--1379}.
\bibitem[{Pannocchia and Rawlings(2003)}]{2003PannocchiaRawlings}
\bibinfo{author}{G.~Pannocchia}, \bibinfo{author}{J.~B. Rawlings},
\newblock \bibinfo{title}{Disturbance models for offset-free model-predictive
  control},
\newblock \bibinfo{journal}{AIChE Journal} \bibinfo{volume}{49}
  (\bibinfo{year}{2003}) \bibinfo{pages}{426--437}.
\bibitem[{{SOBEK}(2000)}]{manual_sobek}
\bibinfo{author}{{SOBEK}}, \bibinfo{title}{Manual and technical reference},
  \bibinfo{type}{Technical Report}, WL$\vert$Delft Hydraulics, Delft, The
  Netherlands, \bibinfo{year}{2000}.
\bibitem[{Isapoor et~al.(2011)Isapoor, Montazar, van Overloop, and van~de
  Giesen}]{Isapoor2011}
\bibinfo{author}{S.~Isapoor}, \bibinfo{author}{A.~Montazar},
  \bibinfo{author}{P.~J. van Overloop}, \bibinfo{author}{N.~van~de Giesen},
\newblock \bibinfo{title}{Designing and evaluating control systems of the {D}ez
  main canal},
\newblock \bibinfo{journal}{Irrigation and Drainage} \bibinfo{volume}{60}
  (\bibinfo{year}{2011}) \bibinfo{pages}{70--79}.
\bibitem[{Hashemy et~al.(2013)Hashemy, Monem, Maestre, and
  Van~Overloop}]{HashemyEtAl2013}
\bibinfo{author}{S.~Hashemy}, \bibinfo{author}{M.~Monem},
  \bibinfo{author}{J.~Maestre}, \bibinfo{author}{P.~Van~Overloop},
\newblock \bibinfo{title}{Application of an in-line storage strategy to improve
  the operational performance of main irrigation canals using model predictive
  control},
\newblock \bibinfo{journal}{Journal of Irrigation and Drainage Engineering}
  \bibinfo{volume}{139} (\bibinfo{year}{2013}) \bibinfo{pages}{635--644}.
\bibitem[{Hashemy and Van~Overloop(2013)}]{HashemyVanOverloop2013}
\bibinfo{author}{S.~Hashemy}, \bibinfo{author}{P.~Van~Overloop},
\newblock \bibinfo{title}{Applying decentralized water level difference control
  for operation of the dez main canal under water shortage},
\newblock \bibinfo{journal}{Journal of Irrigation and Drainage Engineering}
  (\bibinfo{year}{2013}).
\bibitem[{{van Overloop} et~al.(2010){van Overloop}, Negenborn, {De Schutter},
  and Giesen}]{vanOverloop2010}
\bibinfo{author}{P.~J. {van Overloop}}, \bibinfo{author}{R.~R. Negenborn},
  \bibinfo{author}{B.~{De Schutter}}, \bibinfo{author}{N.~C. Giesen},
\newblock \bibinfo{title}{Predictive control for national water flow
  optimization in the netherlands},
\newblock in: \bibinfo{editor}{R.~R. Negenborn}, \bibinfo{editor}{Z.~Lukszo},
  \bibinfo{editor}{H.~Hellendoorn} (Eds.), \bibinfo{booktitle}{Intelligent
  Infrastructures}, volume~\bibinfo{volume}{42} of
  \textit{\bibinfo{series}{Intelligent Systems, Control and Automation: Science
  and Engineering}}, \bibinfo{publisher}{Springer Netherlands},
  \bibinfo{year}{2010}, pp. \bibinfo{pages}{439--461}.
\bibitem[{Clemmens(2012)}]{Clemmens2012}
\bibinfo{author}{A.~Clemmens},
\newblock \bibinfo{title}{Water-level difference controller for main canals},
\newblock \bibinfo{journal}{Journal of Irrigation and Drainage Engineering}
  \bibinfo{volume}{138} (\bibinfo{year}{2012}) \bibinfo{pages}{1--8}.
\bibitem[{van Overloop et~al.(2008)van Overloop, Weijs, and
  Dijkstra}]{Overloop2008a}
\bibinfo{author}{P.~J. van Overloop}, \bibinfo{author}{S.~Weijs},
  \bibinfo{author}{S.~Dijkstra},
\newblock \bibinfo{title}{Multiple model predictive control on a drainage canal
  system},
\newblock \bibinfo{journal}{Control Engineering Practice} \bibinfo{volume}{16}
  (\bibinfo{year}{2008}) \bibinfo{pages}{531 -- 540}.
\bibitem[{{van Overloop} et~al.(2005){van Overloop}, Schuurmans, Brouwer, and
  Burt}]{vanOverloop2005}
\bibinfo{author}{P.~J. {van Overloop}}, \bibinfo{author}{J.~Schuurmans},
  \bibinfo{author}{R.~Brouwer}, \bibinfo{author}{C.~Burt},
\newblock \bibinfo{title}{Multiple-model optimization of proportional integral
  controllers on canals},
\newblock \bibinfo{journal}{Journal of Irrigation and Drainage Engineering}
  \bibinfo{volume}{131} (\bibinfo{year}{2005}) \bibinfo{pages}{190--196}.
\bibitem[{Conte et~al.(2012)Conte, Voellmy, Zeilinger, Morari, and
  Jones}]{ConteEtAl2012_ACC}
\bibinfo{author}{C.~Conte}, \bibinfo{author}{N.~Voellmy},
  \bibinfo{author}{M.~Zeilinger}, \bibinfo{author}{M.~Morari},
  \bibinfo{author}{C.~Jones},
\newblock \bibinfo{title}{{Distributed Synthesis and Control of Constrained
  Linear Systems}},
\newblock in: \bibinfo{booktitle}{Proceedings of the 2012 American Control
  Conference}, \bibinfo{address}{Montreal, Canada}, pp.
  \bibinfo{pages}{6017--6022}.
\bibitem[{Conte et~al.(2013)Conte, Zeilinger, Morari, and
  Jones}]{ConteEtAl2013_ECC}
\bibinfo{author}{C.~Conte}, \bibinfo{author}{M.~Zeilinger},
  \bibinfo{author}{M.~Morari}, \bibinfo{author}{C.~Jones},
\newblock \bibinfo{title}{{Robust Distributed Model Predictive Control of
  Linear Systems}},
\newblock in: \bibinfo{booktitle}{European Control Conference},
  \bibinfo{address}{Zurich, Switzerland}, pp. \bibinfo{pages}{2764--2769}.

\end{thebibliography}
\end{document}